# Comments on: "Lyapunov matrices for a class of time delay systems" by V. L. Kharitonov

Murad Abu-Khalaf    Suat Gumussoy
MIT CSAIL, MIT, Cambridge, MA    IEEE Member
`murad@mit.edu`    `suat@gumussoy.net`

February 19, 2018

## Abstract
We prove that an auxiliary two-point boundary value problem presented in *V. L. Kharitonov, Lyapunov matrices for a class of time delay systems, Systems & Control Letters 55 (2006) 610-617* has linearly dependent boundary conditions, and consequently a unique solution does not exist. Therefore, the two-point boundary value problem presented therein fails to be a basis for constructing Lyapunov matrices for the class of time delay systems investigated.

## 1  Introduction

In [1], the author considers a stable linear time-delay system of the form

$$\dot{x}(t) = A_0 x(t) + A_1 x(t-h) + \int_{-h}^{0} G(\theta) x(t+\theta) d\theta, \qquad t \geq 0 \qquad (1)$$

and is interested in

$$U(\tau) \triangleq \int_{0}^{\infty} \Phi^T(t) W \Phi(t+\tau) dt, \qquad \forall \tau \in \mathbb{R} \qquad (2)$$

where $\Phi(t)$ is the fundamental matrix and $U(\cdot)$ is well-defined and is referred to as a Lyapunov matrix. It was shown in both [1] and [2], and references therein, that $U(\cdot)$ is characterized by a dynamic, a symmetric, and an algebraic property

$$\dot{U}(\tau) = U(\tau) A_0 + U(\tau - h) A_1 + \int_{-h}^{0} U(\tau + \theta) G(\theta) d\theta, \qquad \tau \geq 0$$

$$U(-\tau) = U(\tau)^T \qquad (3)$$

$$-W = U(0) A_0 + U(-h) A_1 + \int_{-h}^{0} U(\theta) G(\theta) d\theta + A_0^T U(0) + A_1^T U(h) + \int_{-h}^{0} G(\theta)^T U(-\theta) d\theta$$

Note that $x(t)$ is an $n \times 1$ vector function of time; $G(\cdot)$ and $U(\cdot)$ are $n \times n$ function matrices; $A_0$, $A_1$, and $W = W^T$ are $n \times n$ constant matrices; and $h$ a nonnegative scalar.

To solve for $U(\cdot)$, [1] proposed solving for (3) indirectly by first solving for an auxiliary two-point boundary value problem and use the solution of this auxiliary system to construct a solution to (3). The following concrete example was provided in Section 4 of [1]

$$\dot{x}(t) = A_0 x(t) + A_1 x(t-1) + \int_{-1}^{0} G(\theta) x(t+\theta) d\theta$$

where

$$A_0 = \begin{bmatrix} -1 & 0 \\ 0 & -1 \end{bmatrix}, \quad A_1 = \begin{bmatrix} 0 & 1 \\ -1 & 0 \end{bmatrix}, \quad B_0 = \begin{bmatrix} 0.3 & 0 \\ 0 & 0.3 \end{bmatrix}, \quad B_1 = \begin{bmatrix} 0 & 0.3 \\ -0.3 & 0 \end{bmatrix}$$

$$G(\theta) = \sin(\pi\theta)B_0 + \cos(\pi\theta)B_1$$

and [1] derived based on its methodology the following auxiliary two-point boundary value problem

$$\begin{aligned}
\dot{Z}(\tau) &= Z(\tau)A_0 + V(\tau)A_1 + X_0(\tau)B_0 + X_1(\tau)B_1 \\
\dot{V}(\tau) &= -A_1^T Z(\tau) - A_0^T V(\tau) - B_0^T X_0(\tau) + B_1^T X_1(\tau) \\
\dot{X}_0(\tau) &= -\pi X_1(\tau) \\
\dot{X}_1(\tau) &= Z(\tau) + V(\tau) + \pi X_0(\tau)
\end{aligned} \quad (4)$$

with boundary conditions

$$\begin{aligned}
-W &= Z(0)A_0 + A_0^T Z(0) + V(0)A_1 + A_1^T V^T(0) + X_0(0)B_0 + B_0^T X_0^T(0) + X_1(0)B_1 + B_1^T X_1^T(0) \\
0 &= Z(0) - V(1) \\
0 &= X_0(0) - X_0^T(1) \\
0 &= X_1(0) + X_1^T(1)
\end{aligned} \quad (5)$$

Once $Z(\tau)$ is computed, then $U(\tau)$ can be constructed

$$\begin{aligned}
U(\tau) &= Z(\tau), \quad \tau \geq 0 \\
U(-\tau) &= Z^T(\tau), \quad \tau \geq 0
\end{aligned} \quad (6)$$

The work [1], and similarly [2], does not examine existence and uniqueness of solutions conditions for the auxiliary two-point boundary value problem (4)-(5). Following our experience in not being able to reproduce the numerical results of this example of Section 4 of [1], we provide in what follows a proof showing that the boundary conditions of system (4)-(5) are in fact linearly dependent. Therefore, (4)-(5) does not yield a well-defined solution, and cannot generate a unique $Z(\tau)$ that can be used to construct a unique $U(\tau)$. In particular, we show that the following set of boundary conditions is linearly dependent

$$\begin{aligned}
0 &= Z(0) - V(1) \\
0 &= X_0(0) - X_0^T(1) \\
0 &= X_1(0) + X_1^T(1)
\end{aligned} \quad (7)$$

## 2    Linear Dependence of Boundary Conditions

Rewriting the auxiliary two-point boundary value problem in Kronecker product form [3], equation (4) becomes

$$\begin{bmatrix} \dot{z} \\ \dot{v} \\ \dot{x}_0 \\ \dot{x}_1 \end{bmatrix} = \overbrace{\begin{bmatrix} A_0^T \otimes I_n & A_1^T \otimes I_n & B_0^T \otimes I_n & B_1^T \otimes I_n \\ -I_n \otimes A_1^T & -I_n \otimes A_0^T & -I_n \otimes B_0^T & I_n \otimes B_1^T \\ 0 & 0 & 0 & -\pi I_n \otimes I_n \\ I_n \otimes I_n & I_n \otimes I_n & \pi I_n \otimes I_n & 0 \end{bmatrix}}^{H} \begin{bmatrix} z \\ v \\ x_0 \\ x_1 \end{bmatrix} \quad (8)$$

where $z = vec(Z)$, $v = vec(V)$, $x_0 = vec(X_0)$, and $x_1 = vec(X_1)$ are $n^2 \times 1$ vectors; $vec(A)$ generates a vector by stacking the columns $A$ into a single column; $I_n$ is an identity matrix of size $n \times n$; $\mathbf{0}$ is a $n^2 \times n^2$ matrix of zeros. Note the following properties follow from [3]

$$vec(A^T) = Tvec(A), \quad T = T^T$$
$$vec(ABC) = (C^T \otimes A)vec(B)$$
$$T(A \otimes C^T)Tvec(B) = T(A \otimes C^T)vec(B^T)$$
$$= Tvec(C^T B^T A^T)$$
$$= vec(ABC)$$
$$= (C^T \otimes A)vec(B)$$

Moreover, equation (7) becomes

$$\begin{bmatrix} 0 \\ \vdots \\ 0 \end{bmatrix} = \begin{bmatrix} I_n \otimes I_n & \mathbf{0} & \mathbf{0} & \mathbf{0} \\ \mathbf{0} & \mathbf{0} & I_n \otimes I_n & \mathbf{0} \\ \mathbf{0} & \mathbf{0} & \mathbf{0} & I_n \otimes I_n \end{bmatrix} \begin{bmatrix} z(0) \\ v(0) \\ x_0(0) \\ x_1(0) \end{bmatrix} + \begin{bmatrix} \mathbf{0} & -I_n \otimes I_n & \mathbf{0} & \mathbf{0} \\ \mathbf{0} & \mathbf{0} & -T & \mathbf{0} \\ \mathbf{0} & \mathbf{0} & \mathbf{0} & T \end{bmatrix} \begin{bmatrix} z(1) \\ v(1) \\ x_0(1) \\ x_1(1) \end{bmatrix} \quad (9)$$

Evaluating the following at $\tau = 1$ and plugging in (9)

$$\begin{bmatrix} z(\tau) \\ v(\tau) \\ x_0(\tau) \\ x_1(\tau) \end{bmatrix} = e^{H\tau} \begin{bmatrix} z(0) \\ v(0) \\ x_0(0) \\ x_1(0) \end{bmatrix} \quad (10)$$

results in

$$\begin{bmatrix} 0 \\ \vdots \\ 0 \end{bmatrix} = \left( \begin{bmatrix} I_n \otimes I_n & \mathbf{0} & \mathbf{0} & \mathbf{0} \\ \mathbf{0} & \mathbf{0} & I_n \otimes I_n & \mathbf{0} \\ \mathbf{0} & \mathbf{0} & \mathbf{0} & I_n \otimes I_n \end{bmatrix} + \begin{bmatrix} \mathbf{0} & -I_n \otimes I_n & \mathbf{0} & \mathbf{0} \\ \mathbf{0} & \mathbf{0} & -T & \mathbf{0} \\ \mathbf{0} & \mathbf{0} & \mathbf{0} & T \end{bmatrix} e^{H\tau} \right) \begin{bmatrix} z(0) \\ v(0) \\ x_0(0) \\ x_1(0) \end{bmatrix} \quad (11)$$

which has $3n^2$ rows. Clearly if the rank of the row space of (11) is less than $3n^2$, then the rows are linearly dependent. Without loss of generality, we add to (9) an additional boundary condition $0 = Z(1)^T - V(0)$ – one we initially guess to be in the row space of (9) – and investigate instead the squared matrix

$$\begin{bmatrix} 0 \\ \vdots \\ 0 \end{bmatrix} = \begin{bmatrix} I_n \otimes I_n & \mathbf{0} & \mathbf{0} & \mathbf{0} \\ \mathbf{0} & I_n \otimes I_n & \mathbf{0} & \mathbf{0} \\ \mathbf{0} & \mathbf{0} & I_n \otimes I_n & \mathbf{0} \\ \mathbf{0} & \mathbf{0} & \mathbf{0} & I_n \otimes I_n \end{bmatrix} \begin{bmatrix} z(0) \\ v(0) \\ x_0(0) \\ x_1(0) \end{bmatrix} - \begin{bmatrix} \mathbf{0} & T & \mathbf{0} & \mathbf{0} \\ T & \mathbf{0} & \mathbf{0} & \mathbf{0} \\ \mathbf{0} & \mathbf{0} & T & \mathbf{0} \\ \mathbf{0} & \mathbf{0} & \mathbf{0} & -T \end{bmatrix} \begin{bmatrix} z(1) \\ v(1) \\ x_0(1) \\ x_1(1) \end{bmatrix} \quad (12)$$

which after evaluating (10) at $\tau = 1$ and plugging in (12) we get

$$\begin{bmatrix} 0 \\ \vdots \\ 0 \end{bmatrix} = \left( \overbrace{\begin{bmatrix} I_n \otimes I_n & 0 & 0 & 0 \\ 0 & I_n \otimes I_n & 0 & 0 \\ 0 & 0 & I_n \otimes I_n & 0 \\ 0 & 0 & 0 & I_n \otimes I_n \end{bmatrix}}^{J} - \overbrace{\begin{bmatrix} 0 & T & 0 & 0 \\ T & 0 & 0 & 0 \\ 0 & 0 & T & 0 \\ 0 & 0 & 0 & -T \end{bmatrix}}^{J} e^H \right) \begin{bmatrix} z(0) \\ v(0) \\ x_0(0) \\ x_1(0) \end{bmatrix} \quad (13)$$

$$\begin{bmatrix} 0 \\ \vdots \\ 0 \end{bmatrix} = (I - Je^H) \begin{bmatrix} z(0) \\ v(0) \\ x_0(0) \\ x_1(0) \end{bmatrix} \quad (14)$$

Note that from $T = T^T$, it follows that $J = J^T$ and that $I = JJ^T = J^T J$. Moreover, system (13) has $4n^2$ rows. If the rank of (13) is *less* than $3n^2$, then this implies at least $n^2 + 1$ rows of (13), and similarly (12), can be eliminated. This implies that (11), and similarly (9), must be linearly dependent. This is further investigated in the subsequent Theorem 1 and Corollary 1 which are influenced by an analysis appearing in [4].

**Lemma 1:** $J^T H J = -H$, where $H$ is given by (8) and $J$ by (13).

*Proof*:

$$\begin{aligned} J^T H J &= \begin{bmatrix} 0 & T & 0 & 0 \\ T & 0 & 0 & 0 \\ 0 & 0 & T & 0 \\ 0 & 0 & 0 & -T \end{bmatrix} \begin{bmatrix} A_0^T \otimes I_n & A_1^T \otimes I_n & B_0^T \otimes I_n & B_1^T \otimes I_n \\ -I_n \otimes A_1^T & -I_n \otimes A_0^T & -I_n \otimes B_0^T & I_n \otimes B_1^T \\ 0 & 0 & 0 & -\pi I_n \otimes I_n \\ I_n \otimes I_n & I_n \otimes I_n & \pi I_n \otimes I_n & 0 \end{bmatrix} \begin{bmatrix} 0 & T & 0 & 0 \\ T & 0 & 0 & 0 \\ 0 & 0 & T & 0 \\ 0 & 0 & 0 & -T \end{bmatrix} \\ &= \begin{bmatrix} -T(I_n \otimes A_1^T) & -T(I_n \otimes A_0^T) & -T(I_n \otimes B_0^T) & T(I_n \otimes B_1^T) \\ T(A_0^T \otimes I_n) & T(A_1^T \otimes I_n) & T(B_0^T \otimes I_n) & T(B_1^T \otimes I_n) \\ 0 & 0 & 0 & -\pi T \\ -T & -T & -\pi T & 0 \end{bmatrix} \begin{bmatrix} 0 & T & 0 & 0 \\ T & 0 & 0 & 0 \\ 0 & 0 & T & 0 \\ 0 & 0 & 0 & -T \end{bmatrix} \\ &= \begin{bmatrix} -T(I_n \otimes A_0^T)T & -T(I_n \otimes A_1^T)T & -T(I_n \otimes B_0^T)T & -T(I_n \otimes B_1^T)T \\ T(A_1^T \otimes I_n)T & T(A_0^T \otimes I_n)T & T(B_0^T \otimes I_n)T & -T(B_1^T \otimes I_n)T \\ 0 & 0 & 0 & \pi TT \\ -TT & -TT & -\pi TT & 0 \end{bmatrix} \\ &= \begin{bmatrix} -A_0^T \otimes I_n & -A_1^T \otimes I_n & -B_0^T \otimes I_n & -B_1^T \otimes I_n \\ I_n \otimes A_1^T & I_n \otimes A_0^T & I_n \otimes B_0^T & -I_n \otimes B_1^T \\ 0 & 0 & 0 & \pi TT \\ -TT & -TT & -\pi TT & 0 \end{bmatrix} \\ &= -H \end{aligned} \quad (15)$$

∎

**Theorem 1:** Dimension of the nullspace of $I - Je^H$ in (14) is $2n^2$, i.e. $\dim Null(I - Je^H) = 2n^2$.

*Proof:* Let $\lambda_i$ be an eigenvalue of $H$ with algebraic multiplicity $m_i$ – geometric multiplicity is $\mu_i$ with $\mu_i \leq m_i$ and represents the dimension of associated eigenspace – whose Jordan block form is $\Sigma_i \in \mathbb{R}^{m_i \times m_i}$ and its corresponding *generalized* eigenvectors $V_i \in \mathbb{R}^{4n^2 \times m_i}$. If $\lambda_i$ is an eigenvalue, then so is $-\lambda_i$

$$\begin{aligned} HV_i &= V_i \Sigma_i \\ J^T H J J^T V_i &= J^T V_i \Sigma_i \\ -H J^T V_i &= J^T V_i \Sigma_i \\ H(J^T V_i) &= (J^T V_i)(-\Sigma_i) \end{aligned} \qquad (16)$$

Thus, all the distinct eigenvalues of $H$ are $\{\pm\lambda_1, \pm\lambda_2, \ldots, \pm\lambda_s\}$ where $\mathrm{Re}\,\lambda_i \leq 0$ and $s \leq 2n^2$ each repeated per its algebraic multiplicity. Let the Jordan form corresponding to $\{\pm\lambda_1, \pm\lambda_2, \ldots, \pm\lambda_s\}$ be

$$\begin{bmatrix} \Sigma & \\ & -\Sigma \end{bmatrix} \in \mathbb{R}^{4n^2 \times 4n^2}$$

where

$$\Sigma = \begin{bmatrix} \Sigma_1 & & \\ & \ddots & \\ & & \Sigma_s \end{bmatrix} \in \mathbb{R}^{2n^2 \times 2n^2}$$

corresponds to $\{\lambda_1, \lambda_2, \ldots, \lambda_s\}$ and their respective generalized eigenvectors $V_i \in \mathbb{R}^{4n^2 \times m_i}$ forming

$$V = [V_1 \cdots V_s] \in \mathbb{R}^{4n^2 \times 2n^2}$$

Note that from (16) it follows that

$$H\begin{bmatrix} V & J^T V \end{bmatrix} = \begin{bmatrix} V & J^T V \end{bmatrix} \begin{bmatrix} \Sigma & 0 \\ 0 & -\Sigma \end{bmatrix}$$

By diagonalizing the exponential matrix – in Jordan Block Form – as follows

$$e^{H\tau} = \begin{bmatrix} V & J^T V \end{bmatrix} \begin{bmatrix} e^{\Sigma \tau} & 0 \\ 0 & e^{-\Sigma \tau} \end{bmatrix} \begin{bmatrix} V & J^T V \end{bmatrix}^{-1} \qquad (17)$$

the state trajectories are written in terms of its eigenmotions

$$\begin{bmatrix} z(\tau) \\ v(\tau) \\ x_0(\tau) \\ x_1(\tau) \end{bmatrix} = \begin{bmatrix} V & J^T V \end{bmatrix} \begin{bmatrix} e^{\Sigma \tau} & 0 \\ 0 & e^{-\Sigma \tau} \end{bmatrix} \begin{bmatrix} V & J^T V \end{bmatrix}^{-1} \begin{bmatrix} z(0) \\ v(0) \\ x_0(0) \\ x_1(0) \end{bmatrix}$$

Letting

$$\begin{bmatrix} c_1 \\ c_2 \\ c_3 \\ c_4 \end{bmatrix} = \begin{bmatrix} V & J^T V \end{bmatrix}^{-1} \begin{bmatrix} z(0) \\ v(0) \\ x_0(0) \\ x_1(0) \end{bmatrix}$$

this simplifies to

$$\begin{bmatrix} z(\tau) \\ v(\tau) \\ x_0(\tau) \\ x_1(\tau) \end{bmatrix} = \begin{bmatrix} V & J^T V \end{bmatrix} \begin{bmatrix} e^{\Sigma \tau} & 0 \\ 0 & e^{-\Sigma \tau} \end{bmatrix} \begin{bmatrix} c_1 \\ c_2 \\ c_3 \\ c_4 \end{bmatrix} \qquad (18)$$

To understand how arbitrary coefficients $c_1,\ldots,c_4$ are, evaluate (18) at $\tau=0$ and plug in (14) to get

$$\begin{bmatrix} 0 \\ \vdots \\ 0 \end{bmatrix} = (I - Je^H)\begin{bmatrix} V & J^T V \end{bmatrix} \begin{bmatrix} c_1 \\ c_2 \\ c_3 \\ c_4 \end{bmatrix} \qquad (19)$$

Evaluating (17) at $\tau=1$ and recalling that $J=J^T$, we have

$$\begin{bmatrix} 0 \\ \vdots \\ 0 \end{bmatrix} = \left(I - J\begin{bmatrix} V & J^T V \end{bmatrix}\begin{bmatrix} e^{\Sigma} & 0 \\ 0 & e^{-\Sigma} \end{bmatrix}\begin{bmatrix} V & J^T V \end{bmatrix}^{-1}\right)\begin{bmatrix} V & J^T V \end{bmatrix}\begin{bmatrix} c_1 \\ c_2 \\ c_3 \\ c_4 \end{bmatrix}$$

$$= \left(\begin{bmatrix} V & J^T V \end{bmatrix} - J\begin{bmatrix} V & J^T V \end{bmatrix}\begin{bmatrix} e^{\Sigma} & 0 \\ 0 & e^{-\Sigma} \end{bmatrix}\right)\begin{bmatrix} c_1 \\ c_2 \\ c_3 \\ c_4 \end{bmatrix}$$

$$= V\left(\begin{bmatrix} c_1 \\ c_2 \end{bmatrix} - e^{-\Sigma}\begin{bmatrix} c_3 \\ c_4 \end{bmatrix}\right) + J^T V \begin{bmatrix} c_3 \\ c_4 \end{bmatrix} - JVe^{\Sigma}\begin{bmatrix} c_1 \\ c_2 \end{bmatrix}$$

$$= \begin{bmatrix} V & J^T V \end{bmatrix} \begin{bmatrix} \begin{bmatrix} c_1 \\ c_2 \end{bmatrix} - e^{-\Sigma}\begin{bmatrix} c_3 \\ c_4 \end{bmatrix} \\ \begin{bmatrix} c_3 \\ c_4 \end{bmatrix} - e^{\Sigma}\begin{bmatrix} c_1 \\ c_2 \end{bmatrix} \end{bmatrix}$$

Since $\begin{bmatrix} V & J^T V \end{bmatrix}$ is linearly independent, it follows that

$$\begin{bmatrix} c_3 \\ c_4 \end{bmatrix} = e^{\Sigma}\begin{bmatrix} c_1 \\ c_2 \end{bmatrix}$$

This implies that

$$\begin{bmatrix} z(0) \\ v(0) \\ x_0(0) \\ x_1(0) \end{bmatrix} = \begin{bmatrix} V & J^T V \end{bmatrix} \begin{bmatrix} c_1 \\ c_2 \\ c_3 \\ c_4 \end{bmatrix}$$

$$= \begin{bmatrix} V & J^T V \end{bmatrix} \begin{bmatrix} I_{2n^2 \times 2n^2} \\ e^{\Sigma} \end{bmatrix} \begin{bmatrix} c_1 \\ c_2 \end{bmatrix} \tag{20}$$

and that $c_1$ and $c_2$ are arbitrary thus

$$\dim Range\left( \begin{bmatrix} V & J^T V \end{bmatrix} \begin{bmatrix} I \\ e^{\Sigma} \end{bmatrix} \right) = 2n^2$$

Plugging (20) into (14), it can be verified that

$$\begin{aligned}
(I - Je^H)\begin{bmatrix} V & J^T V \end{bmatrix}\begin{bmatrix} I \\ e^{\Sigma} \end{bmatrix}\begin{bmatrix} c_1 \\ c_2 \end{bmatrix} &= \begin{bmatrix} V & J^T V \end{bmatrix}\begin{bmatrix} I \\ e^{\Sigma} \end{bmatrix}\begin{bmatrix} c_1 \\ c_2 \end{bmatrix} - J\begin{bmatrix} V & J^T V \end{bmatrix}\begin{bmatrix} e^{\Sigma} & 0 \\ 0 & e^{-\Sigma} \end{bmatrix}\begin{bmatrix} I \\ e^{\Sigma} \end{bmatrix}\begin{bmatrix} c_1 \\ c_2 \end{bmatrix} \\
&= \begin{bmatrix} V & J^T V \end{bmatrix}\begin{bmatrix} I \\ e^{\Sigma} \end{bmatrix}\begin{bmatrix} c_1 \\ c_2 \end{bmatrix} - J\begin{bmatrix} V & J^T V \end{bmatrix}\begin{bmatrix} e^{\Sigma} \\ I \end{bmatrix}\begin{bmatrix} c_1 \\ c_2 \end{bmatrix} \\
&= \begin{bmatrix} V & J^T V \end{bmatrix}\begin{bmatrix} I \\ e^{\Sigma} \end{bmatrix}\begin{bmatrix} c_1 \\ c_2 \end{bmatrix} - \begin{bmatrix} V & J^T V \end{bmatrix}\begin{bmatrix} I \\ e^{\Sigma} \end{bmatrix}\begin{bmatrix} c_1 \\ c_2 \end{bmatrix} \\
&= \begin{bmatrix} 0 \\ \vdots \\ 0 \end{bmatrix}
\end{aligned} \tag{21}$$

Since (20) spans a subspace of the nullspace of $I - Je^H$, $Null(I - Je^H)$, this implies $\dim Null(I - Je^H) \geq 2n^2$.

Equation (14) shows that 1 is an eigenvalue of $Je^H$ and thus $\dim Null(I - Je^H)$ equals the dimension of the generalized eigenspace associated with 1. Similarly, -1 is an eigenvalue of $Je^H$ by noting that

$$(-I - Je^H)\begin{bmatrix} V & J^T V \end{bmatrix}\begin{bmatrix} I \\ -e^{\Sigma} \end{bmatrix}\begin{bmatrix} c_1 \\ c_2 \end{bmatrix} = \begin{bmatrix} 0 \\ \vdots \\ 0 \end{bmatrix}$$

and $\dim Null(-I - Je^H)$ equals the dimension of the generalized eigenspace associated with -1. Thus

$$\dim Range\left( \begin{bmatrix} V & J^T V \end{bmatrix} \begin{bmatrix} I \\ -e^{\Sigma} \end{bmatrix} \right) = 2n^2$$

which spans a subspace of $Null(-I - Je^H)$. This implies $\dim Null(-I - Je^H) \geq 2n^2$. Since the dimensions of the generalized eigenspaces of $Je^H$ is $4n^2$, it follows that

$$\dim Null(I - Je^H) + \dim Null(-I - Je^H) \leq 4n^2$$
$$\dim Null(I - Je^H) \leq 4n^2 - \dim Null(-I - Je^H)$$
$$\leq 2n^2$$

Both $\dim Null(I - Je^H) \geq 2n^2$ and $\dim Null(I - Je^H) \leq 2n^2$ imply that $\dim Null(I - Je^H) = 2n^2$.

∎

**Corollary 1:** The null space of the subsystem (11) has a dimension of at least $n^2$, thus the subsystem (9) is linear dependent.

*Proof:* Since $\dim Null(I - Je^H) = 2n^2$ from Theorem 1, and given that (11) is a $3n^2$-equations subsystem of the $4n^2$-equations system (13), then dimension of the null space of the subsystem (11) is at most $2n^2$, and at least $n^2$. This implies at least $n^2 + 1$ rows of (11), and similarly of (9), can be eliminated and thus implies that (9) is linearly dependent.

∎

## 3   Conclusion

In [5], we propose an alternative structure for an auxiliary two-point boundary value problem whose solution is well-defined, and provide a necessary and sufficient condition for the existence and uniqueness of solutions.